\documentclass[aps,twocolumn, floats,preprintnumbers,showpacs,prd]{revtex4}
\usepackage{graphicx}
\usepackage{url}
\usepackage{amsmath}
\usepackage{amsfonts}
\usepackage{amssymb}
\def\beq{\begin{equation}}
\def\eeq{\end{equation}}
\def\beqa{\begin{eqnarray}}
\def\eeqa{\end{eqnarray}}

\def\ltap{\ \raise.3ex\hbox{$<$\kern-.75em\lower1ex\hbox{$\sim$}}\ }
\def\gtap{\ \raise.3ex\hbox{$>$\kern-.75em\lower1ex\hbox{$\sim$}}\ }
\newcommand{\missET}{\slash{\hspace{-2.5mm}E}_T}

\begin{document}
\preprint{SLAC-PUB-14573} 
\preprint{UCI-HEP-TR-2011-20}

\title{Inelastic Dark Matter at the LHC
}

\author{Yang Bai$^{a}$ and Tim M.P. Tait$^{b}$
\\
\vspace{2mm}
${}^{a}$SLAC National Accelerator Laboratory, 2575 Sand Hill Road, Menlo Park, CA 94025, USA, \\
${}^{b}$Department of Physics and Astronomy,
University of California, Irvine, CA 92697, USA
}

\pacs{12.60.-i, 95.35.+d, 14.80.-j}

\begin{abstract}
We perform a model-independent study of inelastic dark matter at the LHC,
concentrating on the parameter space with the mass splitting 
between the excited and ground states of dark matter above a few hundred MeV,
where the direct detection experiments are unlikely to explore. 
The generic signatures of inelastic dark matter at the LHC are displaced pions together with 
a monojet plus missing energy, and can be tested at the 7 TeV LHC.
\end{abstract}
\maketitle

{\it{\textbf{Introduction.}}}
Despite overwhelming evidence from astrophysical observation, 
we still don't know the particle properties of dark matter (DM)
or how it interacts with standard model (SM) particles. 
Dark matter as a weakly interacting massive particle (WIMP) is predicted 
in many extensions of the SM and motivates searches for its direct detection,
where one looks for ambient WIMPs scattering with heavy nuclei.
An implicit assumption in these searches is 
that DM particles can scatter elastically off detector nuclei. 
So far, there are no unambiguous signals for a detection of DM scattering,
leading to strong bounds on WIMPs with weak scale 
interactions~\cite{Aprile:2010um}.

These null results raise severe questions about the viability of the WIMP
paradigm, and motivate theoretical exploration of modifications of the
standard picture.  For example, if dark matter is required to scatter inelastically
into a state heavier than the initial WIMP by $\gtap 1$~MeV, typical WIMPs
in the Milky Way halo will have insufficient energy to upscatter, explaining the
null results of direct detection experiments.  Nonetheless, we shall see
below that inelastic dark matter (iDM) models can be tested at the Large
Hadron Collider (LHC). The iDM models were introduced some time 
ago~\cite{Han:1997wn,Hall:1997ah} to evade constraints on models from null DM direct detection searches, and were recently revisited to reconcile 
the DAMA observation of annual modulation~\cite{Bernabei:2010mq} 
with other null direct searches~\cite{TuckerSmith:2001hy},
provided the mass splitting is below a few hundred keV. 

In this letter, we explore the LHC's capability to identify iDM scenarios
through a model-independent, effective field theory (EFT) 
approach~\cite{Birkedal:2004xn,Goodman:2010yf,Bai:2010hh,Rajaraman:2011wf,Fox:2011fx}.
We discuss general strategies and characterize signatures appropriate for
different parameter space.  
The mass splitting $\Delta$ and coefficients of the various
EFT operators control the lifetime of the excited DM particle,
but it is generically long-lived on collider time scales. 
If the mass splitting is too small, the decay products of the excited state are too soft 
to be observed and the generic signature is a monojet plus missing 
energy~\cite{Goodman:2010yf,Bai:2010hh,Rajaraman:2011wf}. 
However, for mass splittings $\gtap$~1 GeV, iDM results in 
a spectacular signature with displaced pions appearing
on top of the monojet signature.

{\it{\textbf{Operators and interactions.}}}
We assume the dark sector is composed of two 
SM gauge singlet fermions $\chi$ and 
$\chi_*$ with masses $m$ and $m_*$, where $\Delta \equiv m_* - m > 0$
characterizes the splitting between the two states.  We assume that
interactions with the SM are required to contain one $\chi$ and $\chi_*$
each, and (for simplicity) restrict ourselves to interactions
consisting of operators which
involve the up-quark and preserve parity:
\beqa
{\cal O}_1 &=& \frac{ \left[ \bar{u}\,\gamma_\mu\gamma_5\, u\right] \, 
\left[\bar{\chi}_*\,\gamma^\mu\gamma_5\,\chi \right]}{\Lambda^2_1}
\, , \hspace{2mm}
{\cal O}_2 =\frac{ \left[ \bar{u}\,\gamma_5\,{u} \right] \,
\left[ \bar{\chi}_* \,\gamma_5\,\chi \right]}{\Lambda^2_2} \,, \nonumber \\
{\cal O}_3 &=&\frac{\left[ \bar{u}\,{u} \right]\,
\left[\bar{\chi}_*\,\chi \right]}{\Lambda^2_3}
\,,  \hspace{8mm}
{\cal O}_4 =\frac{\left[ \bar{u}\,\gamma_\mu\,{u} \right]\,
\left[ \bar{\chi}_*\,\gamma^\mu\,\chi \right]}{\Lambda^2_4} 
 \, .
\label{eq:operatorquark}
\eeqa
where $\Lambda_i$ parameterizes the strength of each interaction.
The effective Lagrangian consists of the SM plus
kinetic terms for $\chi$ and $\chi_*$, as well as these operators plus their complex conjugates.

We are interested in mass splittings ranging from 100 MeV to several GeV. For $\Delta \ltap 1$~GeV, we can use the chiral Lagrangian to describe how these
interactions lead to $\chi_*$ decaying into $\chi$ plus a number of pions. We focus on non-vanishing terms with the minimum number of pion legs, since higher order interactions are
phase-space suppressed.  The leading terms for each operator are
\beqa
{\cal O}_1 & \rightarrow & \frac{F_\pi}{2 \Lambda^2_1} \:
\left[ \bar{\chi}_*\,\gamma_5 \gamma^\mu \chi \right] \: 
\left( \partial_\mu \pi^0 \right) \,, \nonumber \\
{\cal O}_2 & \rightarrow & \frac{i\,\langle \bar u u\rangle}{ F_\pi \Lambda_2^2} \:
\left[ \bar{\chi}_*\,\gamma_5\,\chi \right] \: \pi^0 \,, \nonumber \\
{\cal O}_3 & \rightarrow &  -\frac{\langle \bar u u\rangle}{2 F^2_\pi \Lambda^2_3}
\: \left[ \bar{\chi}_*\,\chi \right] \left( \pi^0\pi^0+2\pi^+\pi^- \right) 
\,, \nonumber \\
{\cal O}_4 & \rightarrow &
\frac{1}{\Lambda^2_4} \left[ \bar{\chi}_*\,\gamma^\mu\,\chi \right]
\left( \pi^-\partial_\mu\,\pi^+ - \pi^+ \partial_\mu\,\pi^- \right) \,,
\label{eq:chiraltrans}
\eeqa
where $F_\pi = 184$~MeV and $\langle\bar u u\rangle= - (242~\mbox{MeV})^3$ (evaluated at 1 GeV)~\cite{Jamin:2002ev}). For on-shell $\chi$ and $\chi_*$,
${\cal O}_1$ can be rewritten in terms of $m$ and $m_*$, in analogy with
the usual treatment of pion-nucleon interactions.
The parity of the quark bilinears involved dictate that the leading
terms induced by ${\cal O}_{1,2}$ in the chiral Lagrangian
correspond to one pion interactions, whereas ${\cal O}_{3,4}$
correspond to interactions involving at least two pions.  For theories in
which dark matter also couples to down-quarks, the right-hand side of 
Eq.~(\ref{eq:chiraltrans}) for  ${\cal O}_{1,4}$ project onto the iso-spin violating 
terms, whereas
for ${\cal O}_{2,3}$ arise from iso-spin conserving pieces.

{\it{\textbf{Decay of the $\chi_*$.}}}
We concentrate on the regions of parameter space with $\Delta \geq m_{\pi^0}$
(for ${\cal O}_{1, 2}$) or
$\Delta \geq 2 m_{\pi^+}$ (for ${\cal O}_{3, 4}$), for which
decays lead to real pions and result in hard enough
decay products so as to register in LHC detectors.  We deal with the regimes
$\Delta \ltap 1$~GeV (described by the chiral Lagangrian) and 
$\Delta \gtap 1.5$~GeV (described by interactions with quarks) separately.

\begin{figure}[!]
\begin{center}
\hspace*{-0.75cm}
\includegraphics[width=0.48\textwidth]{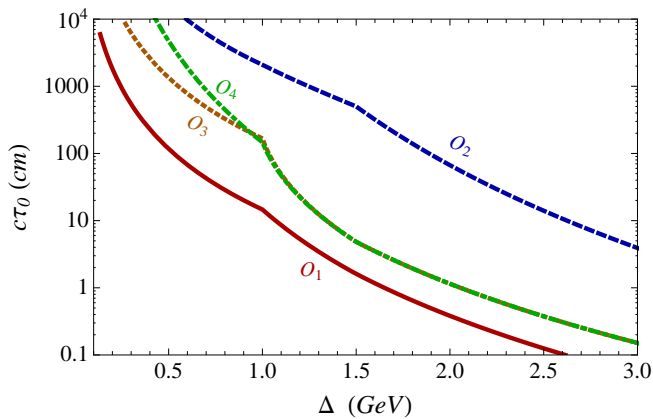}
\caption{Decay lengths of the excited dark matter state at rest as a function of
mass splitting for different operators with $\Lambda_i = 1$~TeV and
a dark matter mass of 5 GeV.}
\label{fig:ctaurest}
\end{center}
\end{figure}

For $\Delta \ltap 1$~GeV,
${\cal O}_1$ and ${\cal O}_2$, lead to two-body decays 
$\chi_* \rightarrow \chi + \pi^0$  with decay widths $\Gamma_i$ ($i$=1,2)
\beq
\alpha_i^2\,
\frac{(\Delta^2 - m^2_{\pi^0}) \sqrt{(\Delta^2 - m^2_{\pi^0}) 
( 4\,\overline{m}^2 - m^2_{\pi^0}) }  }
{16\,\pi\,m^3_{*} } \;,
\eeq
where
$\overline{m} \equiv (m_* + m )/2$,
$\alpha_1 =  F_\pi\,\overline{m} / \Lambda^2_1$,
and $\alpha_2 =  |\langle\bar u u\rangle|/(F_\pi \Lambda^2_2)$.
In the limit $\overline{m} \gg \Delta > m_{\pi^0}$, $\Gamma_1$ is roughly 
independent of the DM mass, and $\Gamma_2 \propto \Delta^3 / \overline{m}^2$.
Operators ${\cal O}_{3,4}$ result in three-body decays of $\chi_*$ to $\chi$ 
plus two pions. 
Neglecting pion masses and in the limit $\overline{m} \gg \Delta$,
\beqa
&&\hspace{-0.5cm}\Gamma_3(\chi_* \rightarrow \chi \pi^+ \pi^-) 
= 2 \Gamma_3(\chi_* \rightarrow \chi \pi^0\pi^0)  
= \frac{\langle \bar u u\rangle^2 \Delta^3}{48\pi^3\,F^4_\pi \Lambda_3^4}
\,, \nonumber \\
&&\hspace{-0.5cm}\Gamma_4(\chi_* \rightarrow \chi \pi^+ \pi^-) = \frac{\Delta^5}{240\pi^3 \Lambda_4^4} \,.
\eeqa

For $\Delta \gtap 1.5$~GeV, the chiral Lagrangian is no longer a
suitable description, and we compute $\chi_* \rightarrow \chi q \bar{q}$,
\beqa
\Gamma(\chi_* \rightarrow \chi u \bar{u})  =  \frac{a_i}{\pi^3}\frac{\Delta^5}{\Lambda_i^4}\,,
\eeqa
where $a_1 = 1/20$, $a_2=\Delta^2/(560\,\overline{m}^2)$, and
$a_3=a_4=1/60$. The decay produces soft jets of hadrons (mostly pions with a small fraction of kaons)
described by the parton shower of QCD.  The intermediate
region of $1~{\rm GeV} \ltap \Delta \ltap 1.5~{\rm GeV}$ is
complicated, and receives contributions from resonances as well as
multi-pion states.  We have approximated the behavior in this region
by requiring the decay length smoothly interpolate between the leading decay
for small $\Delta$ and the soft jet regime of large $\Delta$.

Decay lengths $c \tau_0$ as a function of $\Delta$, for $\Lambda_i = 1$~TeV
and  $\overline{m} = 5$~GeV are shown in Fig.~\ref{fig:ctaurest}.
Different values of $\Lambda_i$ rescale the presented lifetime by 
$(\Lambda_{\rm new} / 1~{\rm TeV} )^4$.  For the chosen parameters,
they vary from 1 cm to 10 m, depending on $\Delta$ and the operator
mediating the decay.  As we shall see below, the most useful production
regime at a hadron collider results in relativistic $\chi_*$ whose
lifetimes in the detector frame are given by
$c\tau = \gamma c \tau_0$, where $\gamma=E_{\chi_*}/m_*$. 
In our detailed LHC calculations below, we 
include this dilation factor on an event-by-event basis. 

{\it{\textbf{Production of iDM particles at the LHC.}}}
At the LHC, the interactions with quarks will result in events containing
one $\chi$ and one $\chi_*$ in the final state.  The 
hadrons from the $\chi_*$ decays are generally too soft to be used as triggers.
However, a ``monojet" process, $p\,p \rightarrow \overline{\chi}_* \,\chi \, j$
(plus the conjugate $\overline{\chi} \chi_*\, j$ process)
containing an additional unflavored jet $j$ radiated from
the initial partons can provide a suitable trigger.
After the $\chi_*$ decay, the final state consists of 
$2 \chi + j + \pi$'s, where because of the long $\chi_*$ lifetime, the $\pi$'s are
produced far from the primary interaction vertex, leading to a signature of a
{\em monojet plus displaced pions}. 

\begin{figure}[!]
\begin{center}
\hspace*{-0.75cm}
\includegraphics[width=0.48\textwidth]{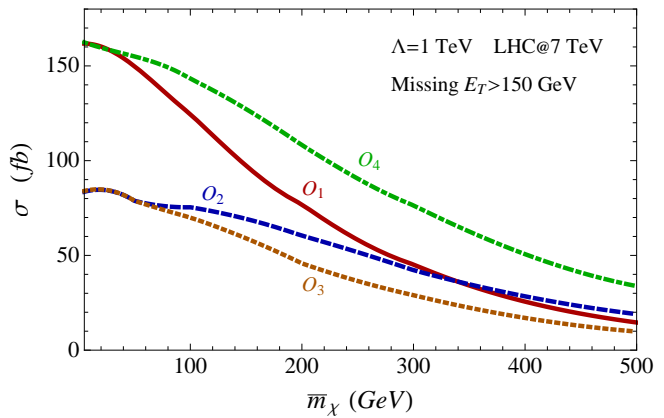}
\caption{Production cross section of $p p \rightarrow \chi_e\, \bar{\chi}_g\, j$ 
at the 7 TeV LHC.}
\label{fig:production}
\end{center}
\end{figure}

Current LHC monojet searches~\cite{Chatrchyan:2011nd,Aad:2011xw} rely on a
missing energy trigger and apply a missing energy cut of 
$\not \hspace*{-0.18cm} E_T > 150$~GeV (at CMS).  
For the iDM signature, one may use the same 
triggers and missing energy cut.  We simulate the expected production rate after
the missing energy cut at the 7 TeV LHC using Madgraph 5~\cite{Alwall:2011uj}
with the CTEQ 6L1~\cite{Pumplin:2002vw} parton distribution functions (PDFs).
The results for the various operators are shown in Fig.~\ref{fig:production}.
For DM masses below around 50 GeV, the rates become independent of the
value of the mass itself, because the $\missET$ cut becomes the limiting
factor for production.

\begin{figure}[!]
\begin{center}
\includegraphics[width=0.48\textwidth]{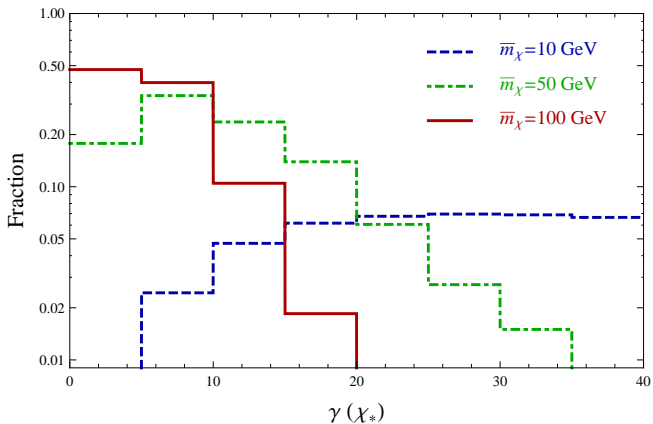}
\caption{The distribution of $\gamma$ (for ${\cal O}_1$) 
for three choices of DM masses after a missing energy cut 
of  $\,\missET \ge 150$~GeV.}
\label{fig:gammafactor}
\end{center}
\end{figure}

In Fig.~\ref{fig:gammafactor}, we show the distribution of the
$\gamma$ factor for the excited WIMP (after cuts)
resulting from production through operator ${\cal O}_1$.  Results from
${\cal O}_{2-4}$ are very similar.  As expected,
the peak of the distribution shifts to larger values for smaller DM masses,
and remains $\sim 10$ for masses as large as 100~GeV.  Comparing
Figs.~\ref{fig:ctaurest} and \ref{fig:gammafactor} reveals that for WIMP masses
around 50 GeV, the lab frame decay length $\gamma c \tau_0$ is around
1 m for a wide range of $\Delta$ and $\Lambda \sim 1$~TeV.

{\it{\textbf{Search strategy and discovery potential.}}}
In addition to the primary jet against which the $\chi \chi_*$ system recoils,
iDM also produces one or more hadrons deep in the
detector from the $\chi_*$ decay.   
For decay lengths on scales of 10 cm to 1 m, the hadrons are likely to deposit most
of their energy in the electromagnetic (ECAL) or hadronic (HCAL) calorimeters.
For the discussion below, we use detailed numbers for the CMS 
detector~\cite{Bayatian:2006zz}, though similar conclusions will also hold for ATLAS.
The hadrons from the $\chi_*$ decay are not typical of QCD jets, since they
contain a much smaller multiplicity of charged particles (particularly when charged
pions are produced after $\chi_*$ traverses the tracker).  Aside from their
relatively long decay length, these features are common between $\chi_*$
and hadronic tau decays (the actual hadron multiplicities from $\chi_*$ decays are different from hadronic taus and depend on the mass splitting).

For a sufficiently loose definition of a hadronic tau, the iDM decays can mimic an
extra jet consistent with the hadronic tau signature.  After being tagged as
a hadronic tau, one can further reduce real tau backgrounds using the decay length.
Our search strategy is thus a monojet together with a tau-tagged jet.  At CMS,
hadronic tau's resulting in visible transverse energy $> 15$~GeV have
a tagging efficiency of $\sim 20\%$, with a fake rate of $0.3\%$~\cite{cmstau}.  
We adopt these numbers as estimates for the (mis)tag rates for the $\chi_*$ decayed jet, though it would be worthwhile to have a proper treatment by the
experimental collaborations using a realistic detector simulation.
Given the
very short $\tau$ lifetime ($c \tau_\tau \sim 100~\mu$m in the $\tau$ rest frame),
we estimate that the background from $W j$, where $W \rightarrow \tau \nu$ will
be essentially entirely removed by a displacement cut $\gtap 10$~cm.

\begin{figure}[!]
\begin{center}
\includegraphics[width=0.48\textwidth]{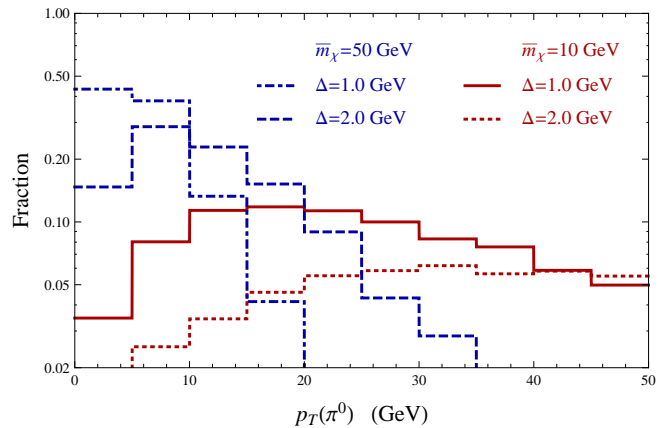}
\caption{The transverse energy of hadrons (resulting from operator ${\cal O}_1$)
for two choices of DM masses and splittings, as indicated.}
\label{fig:ptpion}
\end{center}
\end{figure}

In the limit of the missing $\missET > \overline{m}$ and $\Delta \gg m_{\pi^0}$, 
the total transverse energy of the displaced pions is
$p_T(\pi^0) \sim E_T\,\Delta/\overline{m}_\chi$. 
Lighter $\chi$ masses and larger mass splittings result in more
deposited energy of displaced pions in the calorimeter.  In Fig.~\ref{fig:ptpion}
we show $p_T$ of $\pi^0$ (for ${\cal O}_1$; other operators lead to similar
results) for a few choices of $\overline{m}$ and $\Delta$.
Requiring $p_T(\pi^0) \ge 15$~GeV, the signal efficiency drops
rapidly as the mass varies from $\overline{m} = 50$~GeV to 
$\overline{m} = 10$~GeV.

In order to assess the discovery potential at the LHC, we allow for events
which pass a monojet + ``displaced hadronic tau"
selection.  Thus, we select events with:
$N_{j} = 2$ with $p_T(j_1) > 110$~GeV and $15 < p_T(j_2) < 30$~GeV as well
as  $\missET \ge 150$~GeV.  While in principle one could allow the second
jet (typically from the $\chi_*$ decay) to have a higher $p_T$, in practice for these
choices of $\Delta$ and $p_T(j_1)$, the signal is not very sensitive to the upper
bound on $p_T(j_2)$.  We require the $\chi_*$ decay occur before the
barrel ECAL, 129 cm from the center along the radial 
direction~\cite{Bayatian:2006zz}.  We apply a $20\%$ 
hadronic $\tau$-tagging efficiency
to signal events satisfying these criteria.

We take the backgrounds measured as part of the standard CMS monojet
search based on 36 pb$^{-1}$ \cite{Chatrchyan:2011nd} and apply the
$0.3\%$ hadronic tau mistag rate.  In practice, this is an over-estimate of the
background level because the presented backgrounds include both one and
two jet events; however it can only over-estimate the background, and
suffices for a conservative estimate of the LHC sensitivity.  To estimate
future reach, we rescale the background to a 5~fb$^{-1}$ data set in order 
to determine the future statistical uncertainty
(which is still expectd to be the dominant contribution), thus determining future
$95\%$ limits on a putative signal
at a 7 TeV LHC which has collected 5~fb$^{-1}$.

\begin{figure}[!]
\begin{center}
\includegraphics[width=0.48\textwidth]{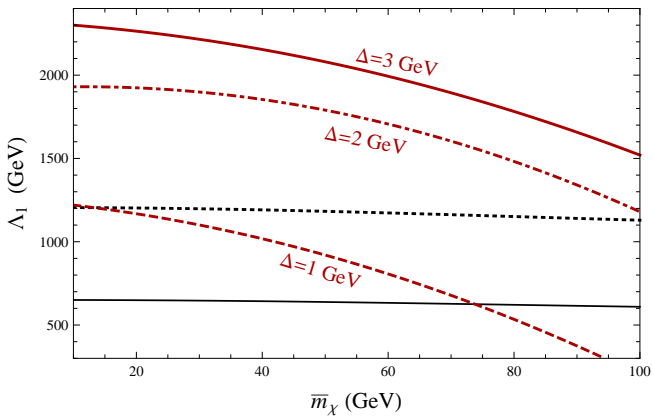}
\caption{The 95\% C.L. exclusion limit on $\Lambda_1$ (for ${\cal O}_1$) as a function of 
$\overline{m}$, assuming 5 fb$^{-1}$ of 7 TeV LHC data, for the three
values of mass splitting $\Delta$ shown.  The approximately straight lines are 
limits derived using monojet only searches at 36 pb$^{-1}$ (solid) 
and 5 fb$^{-1}$ (dotted).}
\label{fig:discPvector}
\end{center}
\end{figure}

\begin{figure}[!]
\begin{center}
\includegraphics[width=0.48\textwidth]{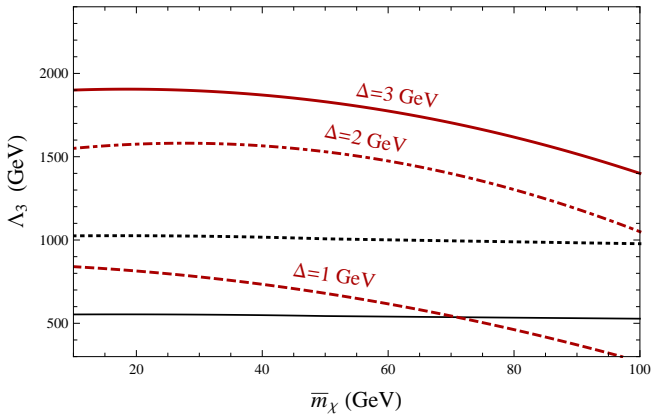}
\caption{The same as Fig.~\ref{fig:discPvector} but the scalar interaction 
${\cal O}_3$.
}
\label{fig:discScalar}
\end{center}
\end{figure}

In Figs.~\ref{fig:discPvector} and \ref{fig:discScalar} we show the expected reach
in terms of the bound on the strength of the contact interaction $\Lambda_i$ for
the axial-vector interaction ${\cal O}_1$ and scalar interaction ${\cal O}_3$,
as a function of $\overline{m}$ and three choices of $\Delta$, for the
LHC operating at 7 TeV and having collected 5 fb$^{-1}$.  Also shown for
reference are the limits from standard monojet searches from the existing
36 pb$^{-1}$ and projected with 5 fb$^{-1}$.  For mass splittings $\gtap 1$~GeV,
limits from the displaced pion search provide more stringent limits on iDM than the
standard monojet search.  For smaller mass splittings, the visible $\chi_*$
decay products have difficulty passing the $p_T \geq 15$ GeV cut.
We find that limits on ${\cal O}_1$ and ${\cal O}_3$ are comparable.  We expect
that the vector operator ${\cal O}_4$ will end up with similar prospects, whereas
the pseudo-scalar ${\cal O}_2$ somewhat worse prospects, because it leads
to a longer $\chi_*$ lifetime, with decays typically happening outside of the detector.
In that case, the standard monojet search will probably better probe the iDM model.

{\it{\textbf{Conclusions.}}}
In conclusion, we have examined iDM models whose larger splitting precludes their direct detection (at tree level -- scattering at one loop level is nonetheless possible,
 but will occur with negligible rate for $\Lambda \sim 1$~TeV). We find that the LHC has the ability to search for such models for a wide variety of masses and splittings, though
a novel signature involving a monojet plus extra hadrons which are reminiscent
of hadronic tau decays, but appear deep in the detector.  We hope these initial
promising results inspire more detailed analyses by the experimental collaborations.

\vspace{3mm}
{\it{\textbf{Acknowledgements.}}}
We thank G.~Landsberg, M.~Peskin and J.~Wacker
for useful discussion. Part of this work was completed at
the Aspen Center for Physics, supported in part by the NSF
under Grant No. 1066293. SLAC is operated by
Stanford University for the US Department of Energy under contract
DE-AC02-76SF00515. TMPT acknowledges the
hospitality of the SLAC theory group, and is supported in part by NSF
grant PHY-0970171.

\end{document}